\documentclass[12pt]{article}
\usepackage{graphicx}
\usepackage{amsmath}
\usepackage{amssymb}
\usepackage{amsfonts}
\usepackage{booktabs}
\usepackage[margin=1in]{geometry}
\usepackage{url}
\begin{document}

\title{ChargeFlow: Flow-Matching Refinement of Charge-Conditioned Electron Densities}

\author{Tri Minh Nguyen, Sherif Abdulkader Tawfik, Truyen Tran, Svetha Venkatesh\\[0.25em]
\normalsize Deakin University, Geelong, VIC 3216, Australia}
\date{}

\maketitle

\begin{abstract}
Accurate charge densities are central to electronic-structure theory, but computing charge-state-dependent densities with density functional theory remains too expensive for large-scale screening and defect workflows. We present ChargeFlow, a flow-matching refinement model that transforms a charge-conditioned superposition of atomic densities into the corresponding DFT electron density on the native periodic real-space grid using a 3D U-Net velocity field. Trained on 9,502 charged Materials Project-derived calculations and evaluated on an external 1,671-structure benchmark spanning perovskites, charged defects, diamond defects, metal-organic frameworks, and organic crystals, ChargeFlow is not uniformly best on every in-distribution class but is strongest on problems dominated by nonlocal charge redistribution and charge-state extrapolation, improving deformation-density error from 3.62\% to 3.21\% and charge-response cosine similarity from 0.571 to 0.655 relative to a ResNet baseline. The predicted densities remain chemically useful under downstream analysis, yielding successful Bader partitioning on all 1,671 benchmark structures and high-fidelity electrostatic potentials, which positions flow matching as a practical density-refinement strategy for charged materials.
\end{abstract}

\section{Introduction}

The electron charge density, $\rho(\mathbf{r})$, is a cornerstone of modern quantum chemistry and computational materials science. According to the Hohenberg-Kohn theorems, this scalar field in three-dimensional space contains all information about the ground state of a many-electron system \cite{hohenberg1964}. As such, accurate knowledge of $\rho(\mathbf{r})$ enables the prediction of a vast range of material properties, from total energy and atomic forces to electronic band structures and chemical reactivity.

Density Functional Theory (DFT) is the predominant computational method used to determine $\rho(\mathbf{r})$ \cite{kohn1965}. Despite its widespread success, DFT's computational expense, which typically scales as the cube of the number of electrons ($O(N^3)$), presents a significant bottleneck for high-throughput materials screening and the simulation of large, complex systems.

To overcome this limitation, machine learning (ML) has been increasingly employed to develop surrogate models that map atomic structures directly to their quantum mechanical properties, bypassing the need for solving the Kohn-Sham equations \cite{butler2018machine}. Numerous approaches have been developed to predict total energies and forces \cite{behler2007generalized, schutt2017schnet}, and more recently, efforts have focused on predicting the electron density itself across both molecules and materials \cite{brockherde2017bypassing, grisafi2019transferable, chandrasekaran2019solving, gong2019predicting, jorgensen2022equivariant, koker2024higher, lee2024convolutional, li2025image, tawfik2025predicting}. Prior work spans transferable density models based on basis expansions and equivariant message passing, grid-point predictors for periodic materials, and convolutional refinement schemes that learn the correction from an atomic-density guess to the self-consistent density \cite{gong2019predicting, koker2024higher, lee2024convolutional, li2025image}. The growing availability of representation-independent crystalline charge-density resources \cite{shen2022representation} and emerging ML models for electron-density response in real space \cite{feng2025efficient} further motivate methods that capture not only static density accuracy but also physically meaningful charge redistribution. Despite this progress, existing density-prediction models can still face challenges in accurately capturing the subtle, non-local quantum effects that govern charge distribution, especially in charged or defective systems \cite{tawfik2025predicting}.

In this work, we reformulate charge-conditioned electron-density prediction as a generative refinement problem. We introduce ChargeFlow, a model based on Continuous Normalizing Flows (CNFs) \cite{chen2018neural} trained with the flow-matching objective \cite{lipman2022flow}, to map a charge-specific superposition of atomic densities (SAD) onto the corresponding self-consistent DFT density. The emphasis of this manuscript is methodological rather than benchmark breadth: we ask whether flow matching can learn a stable refinement map on native periodic grids, improve physically meaningful density-derived quantities beyond pointwise error, and extrapolate more smoothly across charge states than regression baselines.

ChargeFlow uses a 3D convolutional U-Net to parameterize the velocity field of the probability flow, with the charge state encoded implicitly through the charge-specific SAD input. This formulation naturally respects periodicity and makes the refinement target chemically interpretable: the model must transform an atomic reference density into the corresponding bonded, charge-redistributed DFT density. We evaluate the method through external benchmark performance and through observables that are especially relevant to electronic-structure applications, including Bader charges, electrostatic potentials, deformation densities, and charge-response functions.

\section{Methods}

\subsection{Continuous Normalizing Flows with Flow Matching}
We model the electron density, represented as a tensor on a 3D real-space grid, using a continuous normalizing flow (CNF). A CNF defines a mapping from a source distribution $p_0$ to a target distribution $p_1$ through the solution of an ordinary differential equation (ODE):
\begin{equation}
    \frac{d\mathbf{z}_t}{dt} = v_t(\mathbf{z}_t, c), \quad \mathbf{z}_0 \sim p_0,
\end{equation}
where $\mathbf{z}_t$ is the state of the density grid at a virtual time $t \in [0, 1]$, and $v_t$ is a time-dependent velocity field parameterized by a neural network, conditioned on structural information $c$.

We employ the flow matching objective \cite{lipman2022flow}, which trains $v_t$ by regressing it against a pre-defined target velocity field $u_t$. Specifically, we use the Conditional Optimal Transport (OT) path \cite{tong2023improving} between a source density grid $\mathbf{x}_0$ and a target density grid $\mathbf{x}_1$. For a given time $t$, the interpolated state and target velocity are:
\begin{align}
    \mathbf{x}_t &= t\mathbf{x}_1 + (1-t)\mathbf{x}_0, \\
    u_t &= \mathbf{x}_1 - \mathbf{x}_0.
\end{align}
The model is trained to predict this velocity field with a mean-squared error objective. At inference, the final density is generated by starting from an initial grid $\mathbf{z}_0$ and numerically integrating the learned ODE from $t=0$ to $t=1$.

In the direct refinement formulation adopted in this work, the source distribution is the low-fidelity electron density $\mathbf{x}_0$ constructed from a superposition of isolated atomic densities (SAD), and the target $\mathbf{x}_1$ is the ground-truth DFT density. The flow thus learns to continuously transform the SAD into the DFT density. At inference, $\mathbf{z}_0$ is set directly to the SAD of the query structure, and the ODE is integrated forward to produce the predicted density.

\subsection{Model Architecture}
The velocity field $v_t$ is parameterized by a 3D U-Net adapted from guided-diffusion architectures for volumetric data. The network operates directly on the interpolated density grid $\mathbf{x}_t$ and predicts a single-channel velocity field on the native VASP grid. To accommodate variable grid sizes, we apply a learnable 3D downsampling block before the encoder and a matching upsampling block at the output. The production model uses four resolution levels with base channel count $C=16$, channel multipliers $[1, 2, 2, 4]$, circular padding to preserve periodic boundary conditions, and nearest-neighbour upsampling in the decoder.

Time conditioning is introduced through sinusoidal embeddings followed by a two-layer MLP, and the resulting conditioning vector is injected into residual blocks through feature-wise linear modulation (FiLM) \cite{perez2018film}. Self-attention is applied only at the coarsest resolution level to capture long-range charge redistribution at manageable cost. No separate charge-label embedding is used: the target charge state is encoded implicitly through the charge-specific SAD input and therefore through $\mathbf{x}_t$. Additional block-level architecture details are collected in Section~S2 of the Supporting Information.

\subsection{Training Procedure and Hybrid Objective}
In the direct refinement mode used for the production model, the flow learns to transform the SAD $\mathbf{x}_0$ directly into the DFT target density $\mathbf{x}_1$. At each training step, a time $t$ is sampled uniformly from $[0, 1]$, and the Conditional OT interpolant $\mathbf{x}_t = t\mathbf{x}_1 + (1-t)\mathbf{x}_0$ and target velocity $u_t = \mathbf{x}_1 - \mathbf{x}_0$ are computed.

We employ a hybrid loss function that combines the flow matching objective with an auxiliary density-level loss:
\begin{equation}
    \mathcal{L} = \mathcal{L}_{\text{FM}} + \alpha \cdot \mathcal{L}_{\text{NormMAE}}.
\end{equation}
The primary term $\mathcal{L}_{\text{FM}}$ is the mean-squared error between the predicted and target velocity fields:
\begin{equation}
    \mathcal{L}_{\text{FM}} = \mathbb{E}_{t, \mathbf{x}_0, \mathbf{x}_1} \left[ \| v_t(\mathbf{x}_t, c) - u_t \|^2_2 \right].
\end{equation}
The auxiliary term $\mathcal{L}_{\text{NormMAE}}$ is a normalized mean absolute error (NormMAE) on the final predicted density, approximated via single-step Euler integration from the current state:
\begin{equation}
    \mathcal{L}_{\text{NormMAE}} = \mathbb{E} \left[ \frac{\sum_i |\hat{\mathbf{x}}_1(\mathbf{r}_i) - \mathbf{x}_1(\mathbf{r}_i)|}{N_e} \right],
\end{equation}
where $\hat{\mathbf{x}}_1 = \mathbf{x}_t + (1-t) \, v_t(\mathbf{x}_t, c)$ is the single-step density estimate, the sum runs over all grid voxels, and $N_e = \sum_i \mathbf{x}_1(\mathbf{r}_i)$ is the total electron count computed as the sum of the target density over the grid. This term directly penalizes errors in the integrated density and ensures that the learned velocity field produces physically accurate charge distributions. The auxiliary loss weight is set to $\alpha = 2.0$.

\subsection{Data Augmentation}
During training, random 90$^\circ$ rotations about the three Cartesian axes are applied to both the SAD and DFT density grids. The same rotation is applied consistently to both grids in each training pair. Because the lattice vectors define a canonical orientation for periodic systems, this augmentation is intended as a robustness aid rather than as a substitute for full rotational equivariance.

\subsection{Inference}
At inference time, the ODE is integrated from $t=0$ (starting from the SAD) to $t=1$ using the Heun method (second-order) with 50 function evaluations. An exponential moving average (EMA) of the model weights (decay rate 0.999 with linear warmup) is maintained during training and used for inference.

\subsection{Dataset and Training Details}
We use the released Materials Project charged-density corpus (\texttt{subMP\_12k\_charged}) as the main training source. The full corpus contains 11,878 charged density calculations derived from 4,145 parent MP structures. The nonzero charge states are approximately balanced, with 1,245--1,333 examples for each of $Q \in \{-3,-2,-1,+1,+2,+3\}$ and 4,104 neutral examples. After filtering invalid grid-size pairs, 3,780 parent structures contribute three charge states, 173 contribute two, and 192 contribute one. For development, the released list-generation script performs a random 80/20 split over charged instances, yielding 9,502 training examples from 4,065 parent MP structures and 2,376 internal development hold-out examples from 1,952 parent MP structures. Because this split is performed at the charged-instance level, 1,872 parent MP identifiers appear in both partitions. We therefore treat it as a development split rather than as an independent estimate of generalization and reserve all distribution-shift claims for the external benchmark described below.

Ground truth electron densities were computed using DFT with the VASP code \cite{kresse1996efficient} and the PBE exchange-correlation functional \cite{perdew1996generalized}. A plane-wave energy cutoff of 520~eV was used with $\Gamma$-centered $k$-point grids. For charged systems, the total number of electrons was adjusted via the NELECT tag in VASP, and a uniform compensating background charge was automatically applied to maintain charge neutrality of the periodic cell, following standard practice for charged defect calculations. Structures were relaxed until forces were below 0.01~eV/\AA. Electronic self-consistency was converged to $10^{-6}$~eV, and calculations that failed to reach SCF convergence within 200 electronic steps were discarded (approximately 3\% of total calculations). The targets stored in the dataset are total electron densities on the native VASP real-space grids; spin density and magnetization are not modeled in this study. The real-space grids vary across structures depending on the unit cell geometry and plane-wave cutoff, and densities are stored in units of e/\AA$^3$.

The periodic test set comprises 1,671 structures spanning eight material classes: perovskites, charged defects, multisite diamond defects, special diamond defects, MOFs, extreme MOFs, organic crystals, and extreme organic crystals. Table~\ref{tab:dataset_breakdown} summarizes both the training split and the external benchmark, including sample counts, unique parent structures, and charge-state coverage. This benchmark is intentionally heterogeneous and class-imbalanced. In particular, the organic-crystal subsets are best interpreted as targeted charge-extrapolation probes rather than broad statistical benchmarks. Among the benchmark subsets with explicit MP identifiers (perovskites and organic crystals), 41 of 43 parent materials are absent from the MP training list; only \texttt{mp-5878} and \texttt{mp-8402} overlap. The remaining classes come from separate defect, diamond, and MOF collections.

\begin{table}[t]
\centering
\scriptsize
\caption{Dataset breakdown for the Materials Project charged-density corpus and the external periodic benchmark. ``Unique parents'' are counted after removing the \texttt{\_charge\_Q} suffix from the structure identifier.}
\label{tab:dataset_breakdown}
\begin{tabular}{lrrl}
\toprule
Split / class & Samples & Unique parents & Charge states \\
\midrule
MP charged-density train & 9502 & 4065 & $\{-3,-2,-1,0,1,2,3\}$ \\
MP charged-density dev hold-out & 2376 & 1952 & $\{-3,-2,-1,0,1,2,3\}$ \\
\midrule
Perovskites & 159 & 40 & $\{-2,-1,1,2\}$ \\
Charged defects & 1149 & 588 & $\{-3,-2,-1,1,2,3\}$ \\
Multisite diamond defects & 54 & 18 & $\{-2,-1,1,2\}$ \\
Special diamond defects & 42 & 14 & $\{-2,-1,1,2\}$ \\
MOFs & 120 & 45 & $\{-2,-1,1,2\}$ \\
Extreme MOFs & 123 & 49 & $\{-20,-10,10,20\}$ \\
Organic crystals & 12 & 3 & $\{-2,-1,1,2\}$ \\
Extreme organic crystals & 12 & 3 & $\{-10,-5,5,10\}$ \\
\bottomrule
\end{tabular}
\end{table}

The model was trained using distributed data-parallelism across 8 NVIDIA V100 GPUs on a single node, with a per-GPU batch size of 1 (effective batch size 8). We used the AdamW optimizer \cite{loshchilov2019decoupled} with $\beta_1 = 0.9$, $\beta_2 = 0.95$, an initial learning rate of $5 \times 10^{-4}$, and linear learning rate decay to near-zero over the course of training. Training was run for 10,000 epochs. A dropout rate of 0.1 was applied in all residual blocks. An exponential moving average (EMA) of the model weights was maintained with a decay rate of 0.999 and an initial warmup phase; the EMA weights were used for all inference and evaluation. All results reported here use the same architecture family and a fixed hyperparameter set across benchmark classes, without class-specific tuning. The external periodic benchmark was used only for final evaluation and was not part of the automated training or checkpoint-selection pipeline.

\section{Results and Discussion}

\subsection{External Benchmark Accuracy}
We evaluated the accuracy of ChargeFlow on the external 1,671-structure periodic test set described above. The benchmark spans eight material classes with very different system sizes, charge ranges, and sample counts and therefore probes both interpolation and extrapolation behavior. We use the mean absolute error percentage ($\varepsilon_{MAE}$) as our primary metric:
\begin{equation}
    \varepsilon_{MAE} = \frac{\sum_i |\rho_{\text{DFT}}(\mathbf{r}_i) - \rho_{\text{ML}}(\mathbf{r}_i)| \, \Delta V}{\sum_i |\rho_{\text{DFT}}(\mathbf{r}_i)| \, \Delta V} \times 100,
\end{equation}
where the sums run over all voxels in the real-space grid, $\Delta V = \Omega / N_{\text{grid}}$ is the voxel volume, $\Omega$ is the unit cell volume, and $N_{\text{grid}}$ is the total number of grid points. Densities are stored in units of e/\AA$^3$. Because $\Delta V$ cancels in the ratio, $\varepsilon_{MAE}$ is independent of cell size and grid resolution, enabling meaningful comparison across structures with different lattice parameters.

The results, summarized in Table~\ref{tab:mae_results}, compare ChargeFlow against four baselines: a ResNet regression model, a U-Net regression model, CDeepDFT \cite{tawfik2025predicting}, and ChargeFlowL2, a variant trained with only the flow-matching loss $\mathcal{L}_{\text{FM}}$ (i.e., $\alpha = 0$) to isolate the contribution of the auxiliary density-level term. All models were trained on the same charged Materials Project-derived corpus and evaluated on the same external periodic benchmark using a common prediction and post-processing pipeline. Representative qualitative slices are provided in the Supporting Information.

ChargeFlow is not uniformly best on every class. On material classes that are relatively well represented by the training distribution, such as perovskites and charged defects, several regression baselines achieve lower $\varepsilon_{MAE}$ than ChargeFlow. The strengths of ChargeFlow appear instead on classes that require larger charge-state extrapolation or more intricate charge redistribution. It achieves the lowest $\varepsilon_{MAE}$ on multisite diamond defects (6.15\%), special diamond defects (6.78\%), MOFs (7.56\%), extreme MOFs (7.84\%), organic crystals (7.60\%), and extreme organic crystals (7.52\%). This pattern suggests that the flow-matching formulation is especially useful when a model must refine a physically meaningful initial density under nontrivial distribution shift.

\textbf{Perovskites and Defective Systems:} The model achieves $\varepsilon_{MAE}$ of 5.41\% on perovskite structures. We note that baseline regression models (CDeepDFT: 3.03\%, U-Net: 3.58\%) outperform ChargeFlow on this class, likely because perovskites are well-represented in the training distribution and regression models can more directly fit these in-distribution patterns. On defective diamond structures, however, ChargeFlow achieves the best performance on multisite defects (6.15\%) and special defects (6.78\%), outperforming all baselines. Diamond is a covalently bonded material where defects induce complex, long-range changes in the electron density, and the flow matching framework's ability to model such redistributions is a key strength. On the much larger charged-defect subset, ChargeFlow remains reasonably accurate at 6.46\% but does not match the best ResNet baseline (4.68\%).

\textbf{Electronically Challenging Materials:} We also tested ChargeFlow on two particularly challenging classes of materials. The \textit{Extreme MOFs} dataset was specifically designed to test the model's ability to extrapolate to charge states unseen during training. While the training data included charges from $-3$ to $+3$, this test set contains metal-organic frameworks with large pores and unusual coordination environments subjected to charge levels of $\pm 10$ and $\pm 20$. Despite this severe distributional shift, the model achieves $\varepsilon_{MAE}$ of 7.84\%, the best among all models (CDeepDFT failed to converge on these systems, indicated by ``nan'' in Table~\ref{tab:mae_results}). This result supports the view that ChargeFlow has learned a transferable refinement of charge redistribution rather than simply interpolating between training charge states.

\begin{table}[h!]
\centering
\caption{Mean Absolute Error $\varepsilon_{\text{MAE}}$ (\%) on the external periodic test classes. Sample counts are given in parentheses. \textbf{Bold} indicates the best result per row. ChargeFlowL2 is a variant trained without the auxiliary MAE loss ($\alpha=0$). Extreme MOFs contain $Q \in \{-20, -10, +10, +20\}$; extreme organic crystals contain $Q \in \{-10, -5, +5, +10\}$. ``nan'' indicates that the model failed to produce valid predictions for that test set.}
\label{tab:mae_results}
\begin{tabular}{lccccc}
\hline
\textbf{System Type} & \textbf{ChargeFlow} & ChargeFlowL2 & ResNet & U-Net & CDeepDFT \\
\hline
Perovskites & 5.41 & 7.12 & 3.74 & 3.58 & \textbf{3.03}\\
Charged Defects & 6.46 & 9.00 & \textbf{4.68} & 5.53 & 5.26\\
Multisite Defects & \textbf{6.15} & 6.43 & 8.60 & 8.59 & 9.67\\
Special Defects & \textbf{6.78} & 7.73 & 8.50 & 7.34 & 9.54\\
MOFs & \textbf{7.56} & 11.29 & 7.86 & 8.78 & 10.80\\
Extreme MOFs & \textbf{7.84} & 10.80 & 8.50 & 7.96 & nan\\
Organic Crystals & \textbf{7.60} & 11.60 & 8.30 & 7.71 & 10.94 \\
Extreme Organic Crystals & \textbf{7.52} & 12.73 & 8.39 & 8.83 & nan\\
\hline
\end{tabular}
\end{table}

\subsection{Downstream Validation: Bader Charge Analysis}

The next two subsections assess the absolute physical fidelity of the predicted densities through downstream observables. A first test is whether the predicted density can reproduce Bader charges \cite{henkelman2006fast}. To this end, we performed Bader charge analysis directly on both ChargeFlow and ResNet predicted densities and compared the resulting per-atom charges to those obtained from the ground-truth DFT densities, without any additional VASP self-consistent calculation or density relaxation. Bader analysis partitions the total electron density into atomic volumes using zero-flux surfaces of the charge density gradient, yielding chemically meaningful atomic charges that are sensitive to the local electronic environment. Unlike integrated metrics such as $\varepsilon_{\text{MAE}}$, Bader charges probe whether the predicted density correctly captures the spatial partitioning of electrons among atoms, a prerequisite for reliable predictions of chemical bonding, oxidation states, and charge transfer.

\begin{figure}[h!]
    \centering
    \includegraphics[width=\linewidth]{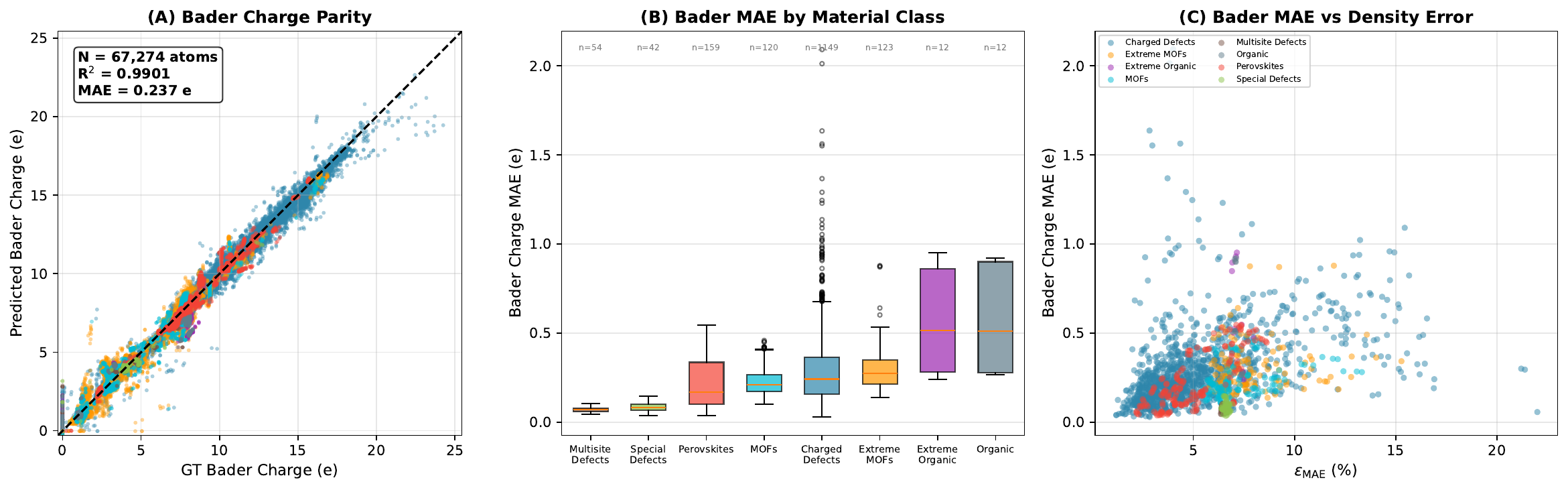}
    \caption{Detailed Bader charge analysis for ChargeFlow across 1,671 periodic materials (67,274 atoms). (A)~Atom-level parity plot of predicted versus ground-truth Bader charges, showing $R^2 = 0.9901$ and MAE~$= 0.237$~e. (B)~Distribution of per-material Bader-charge MAE by material class. (C)~Correlation between electron density error ($\varepsilon_{\text{MAE}}$) and per-material Bader-charge MAE, demonstrating that density accuracy translates to accurate charge partitioning. Table~\ref{tab:downstream_compare} reports the ChargeFlow--ResNet comparison.}
    \label{fig:bader}
\end{figure}

Table~\ref{tab:downstream_compare} summarizes the head-to-head downstream comparison, while Fig.~\ref{fig:bader} retains the detailed per-class ChargeFlow breakdown and the Supporting Information shows the corresponding ResNet analysis. ChargeFlow yields successful Bader partitions for all 1,671 periodic materials (67,274 atoms), with overall atom-level $R^2 = 0.9901$ and atom-level MAE of 0.2369~e. ResNet produces valid Bader outputs for 1,590 of 1,671 materials; the 81 failures are concentrated in charged defects (68), extreme MOFs (8), and MOFs (5). On the 1,590-material common subset where both models produce a valid partition, ChargeFlow retains higher atom-level fidelity than ResNet ($R^2 = 0.9904$ vs.\ 0.9833; atom-level MAE 0.2303~e vs.\ 0.2853~e). These results indicate that ChargeFlow produces more robust and more accurate charge partitioning under downstream post-processing.

The per-atom Bader charge parity plot (Fig.~\ref{fig:bader}A) shows tight clustering along the diagonal across all material classes, with deviations predominantly occurring for atoms with large absolute charges. The breakdown by material class (Fig.~\ref{fig:bader}B) reveals that the model performs best on diamond defect systems, achieving Bader MAEs of 0.071~e for multisite defects and 0.086~e for special defects. These low errors reflect the model's ability to accurately capture the localized charge redistribution induced by vacancies and substitutional impurities in the diamond lattice. Perovskites (0.222~e) and MOFs (0.233~e) also show strong performance, while the organic-crystal subsets yield the largest Bader errors within the periodic benchmark (both approximately 0.56~e), consistent with their small sample size and stronger charge extrapolation regime.

Notably, the extreme MOF test set, which contains charge states of $\pm 10$ and $\pm 20$ far outside the training range of $\pm 3$, achieves a Bader MAE of only 0.306~e. This result further supports the conclusion that ChargeFlow has learned a physically generalizable model of charge redistribution rather than merely interpolating between training charge states. The correlation between density error $\varepsilon_{\text{MAE}}$ and Bader MAE (Fig.~\ref{fig:bader}C) confirms that improvements in density prediction directly translate to more accurate derived properties, validating the use of $\varepsilon_{\text{MAE}}$ as a meaningful proxy for downstream task performance.

\begin{table}[t]
\centering
\caption{Head-to-head downstream comparison on the 1,671-material periodic benchmark. Bader metrics are computed directly on predicted densities. ResNet Bader statistics are reported on the 1,590 structures for which the Bader partition completed successfully; potential metrics are reported on all 1,671 materials.}
\label{tab:downstream_compare}
\begin{tabular}{lcc}
\toprule
Metric & ChargeFlow & ResNet \\
\midrule
Bader successful materials & \textbf{1671 / 1671} & 1590 / 1671 \\
Bader atom-level $R^2$ & \textbf{0.9901} & 0.9833 \\
Bader atom-level MAE (e) & \textbf{0.2369} & 0.2853 \\
Potential mean per-material MAE (eV) & 1.3326 & \textbf{1.1700} \\
Potential mean per-material $R^2$ & \textbf{0.9954} & 0.9928 \\
\bottomrule
\end{tabular}
\end{table}

\subsection{Downstream Validation: Electrostatic Potential}

As a second downstream validation, we assessed whether the predicted electron densities yield accurate electrostatic (Hartree) potentials. The Hartree potential $V_H(\mathbf{r})$ is obtained by solving the Poisson equation $\nabla^2 V_H = -4\pi\rho$ in reciprocal space via FFT:
\begin{equation}
    \tilde{V}_H(\mathbf{G}) = \frac{4\pi\,\tilde{\rho}(\mathbf{G})}{|\mathbf{G}|^2}, \quad \mathbf{G} \neq 0,
\end{equation}
where $\tilde{\rho}(\mathbf{G})$ denotes the Fourier transform of the electron density. The $\mathbf{G} = 0$ component, which sets the absolute potential reference, is excluded from both the predicted and ground-truth potentials. For charged systems, this exclusion corresponds to using the standard convention where the compensating background potential is subtracted. The remaining potential is then aligned by subtracting the spatial mean of each potential before computing the MAE, ensuring that the comparison reflects the spatial variation of the potential rather than an arbitrary offset.

Because the potential depends on the density through a global integral (the Coulomb kernel $1/|\mathbf{r}-\mathbf{r}'|$), errors in the density are amplified non-locally, making this a demanding test of the predicted charge distribution's long-range accuracy.

\begin{figure}[h!]
    \centering
    \includegraphics[width=\linewidth]{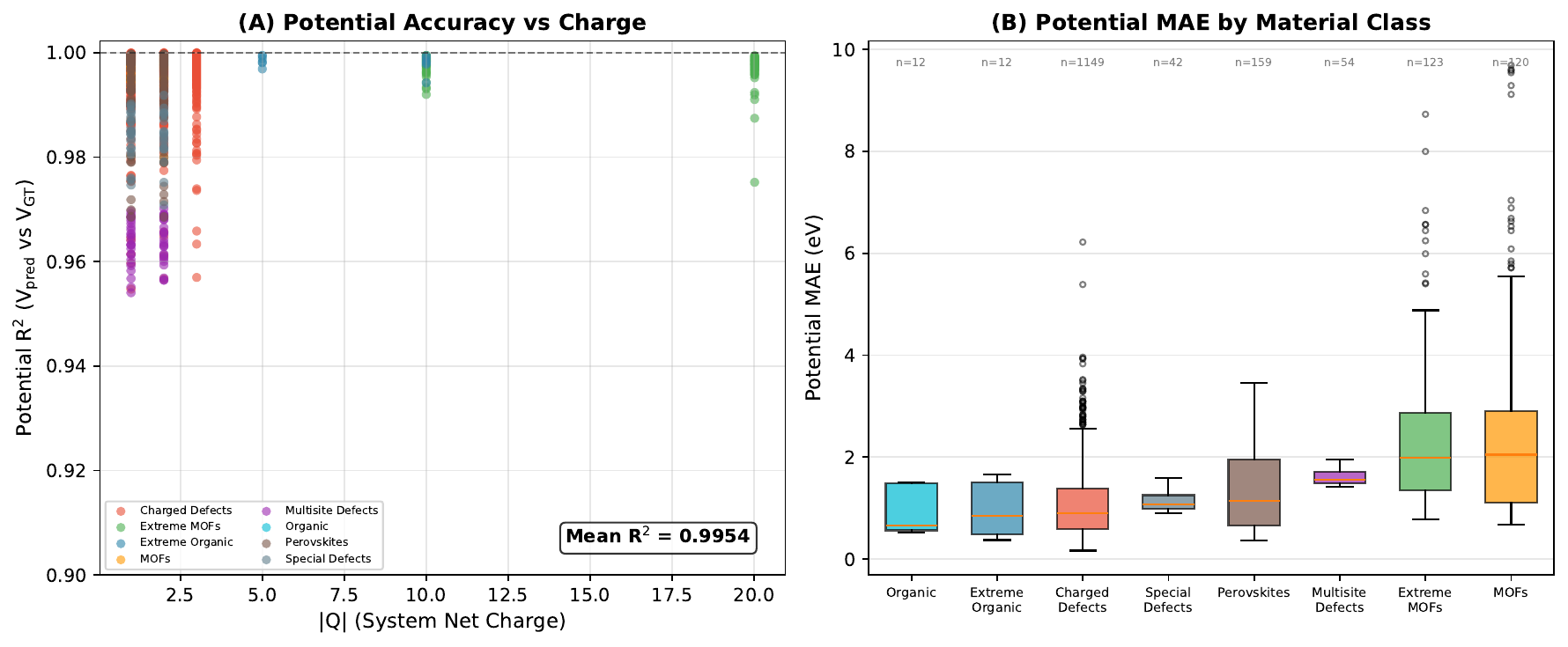}
    \caption{Detailed electrostatic (Hartree) potential analysis for ChargeFlow across 1,671 periodic materials. (A)~Mean per-material potential $R^2$ as a function of system net charge $|Q|$, showing stable accuracy even for extreme charge states (overall mean $R^2 = 0.9954$). (B)~Distribution of per-material potential MAE by material class. Table~\ref{tab:downstream_compare} reports the ChargeFlow--ResNet comparison.}
    \label{fig:potential}
\end{figure}

\begin{figure}[h!]
    \centering
    \includegraphics[width=\linewidth]{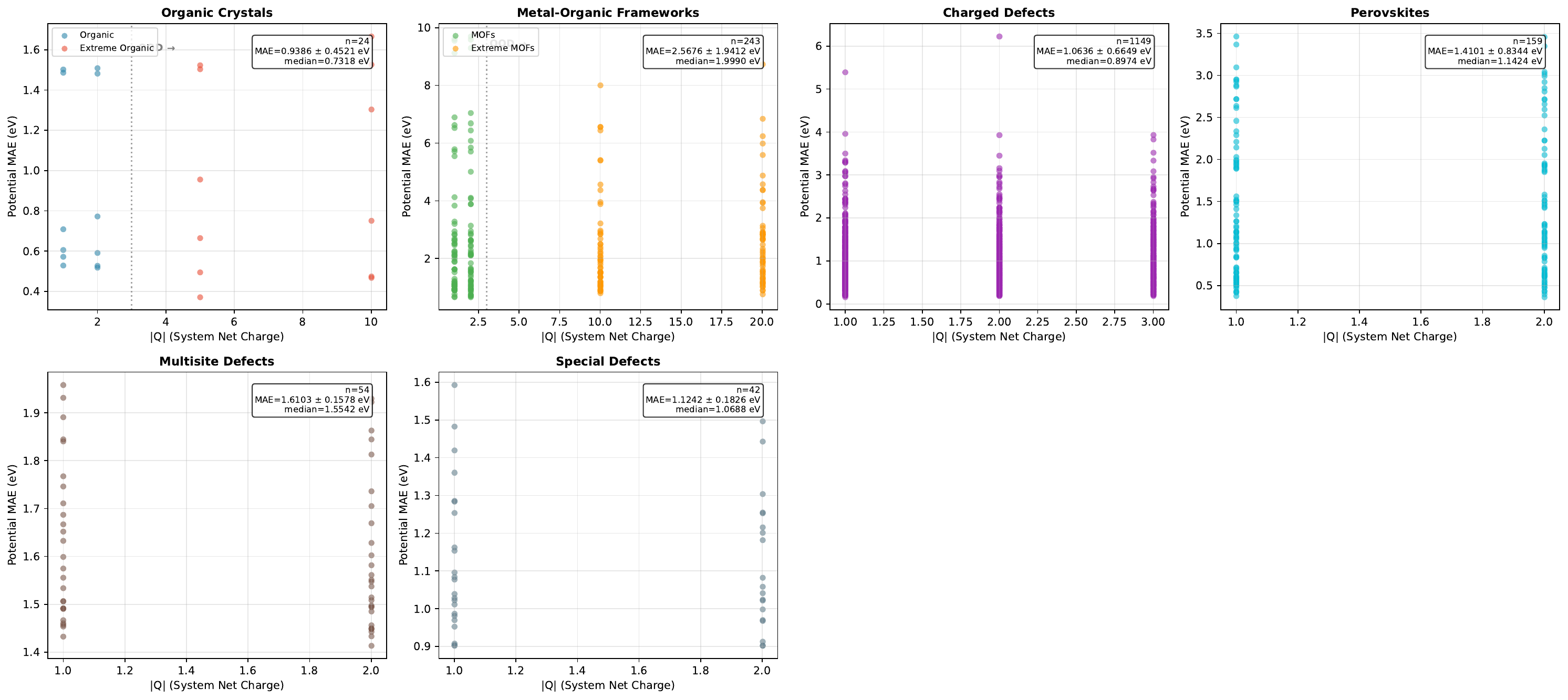}
    \caption{Per-class breakdown of electrostatic potential MAE as a function of system net charge $|Q|$. Each panel corresponds to a different material class, with the number of test structures ($n$), mean MAE, standard deviation, and median indicated. The diamond defect classes exhibit the tightest distributions, while MOFs show the largest variability owing to their complex pore geometries and extreme charge states.}
    \label{fig:potential_testset}
\end{figure}

We computed $V_H$ from both the ChargeFlow- and ResNet-predicted densities, as well as from the ground-truth DFT densities, for all 1,671 periodic test materials. Table~\ref{tab:downstream_compare} reports the overall comparison, while Figs.~\ref{fig:potential} and \ref{fig:potential_testset} retain the detailed ChargeFlow breakdown; the corresponding ResNet plots are provided in the Supporting Information. Both models reproduce the electrostatic potential with high fidelity. ChargeFlow attains a higher mean per-material $R^2$ (0.9954 vs.\ 0.9928), whereas ResNet attains a lower mean per-material MAE on the full benchmark (1.1700 vs.\ 1.3326~eV).

The class-resolved pattern is not uniform. ResNet is more accurate on charged defects, perovskites, and MOF subsets, whereas ChargeFlow is more accurate on the organic and diamond-defect stress tests that emphasize nontrivial charge redistribution. In particular, ChargeFlow lowers the potential MAE from 1.257 to 0.901~eV on organic crystals, from 1.271 to 0.976~eV on extreme organic crystals, from 2.383 to 1.610~eV on multisite diamond defects, and from 1.759 to 1.124~eV on special defects. Averaged across these four charge-redistribution-focused classes, ChargeFlow reduces the mean potential MAE from 1.94 to 1.31~eV and raises the mean potential $R^2$ from 0.942 to 0.978.

Within ChargeFlow itself, the potential accuracy remains remarkably stable across charge magnitudes (Fig.~\ref{fig:potential}A). For systems with $|Q| < 5$, the MAE is 1.23--1.33~eV. Even for extreme charge states of $|Q| \in [10, 20)$ and $|Q| \in [20, 100)$, the MAE increases only modestly to 2.23 and 2.59~eV respectively, while $R^2$ remains above 0.997. This demonstrates that the flow-matching framework produces densities whose long-range electrostatic properties degrade gracefully under severe charge perturbations.

The detailed ChargeFlow breakdown by material class (Fig.~\ref{fig:potential}B) shows that organic crystals and charged diamond defects achieve the lowest potential MAEs (0.90--1.12~eV), followed by perovskites (1.41~eV). MOFs and extreme MOFs exhibit higher errors (2.47--2.66~eV), consistent with their more complex pore geometries and the larger spatial extent over which charge redistribution must be captured.

A more detailed analysis of the potential accuracy per material class is presented in Fig.~\ref{fig:potential_testset}. Among the crystalline systems, organic crystals exhibit the lowest error ($\text{MAE} = 0.94 \pm 0.45$~eV, median~$= 0.73$~eV), followed by charged diamond defects ($1.06 \pm 0.66$~eV) and special defects ($1.12 \pm 0.18$~eV). Perovskites achieve an MAE of $1.41 \pm 0.83$~eV. Notably, the diamond defect classes (multisite and special) display exceptionally tight error distributions, reflecting the model's consistent treatment of vacancy- and substitution-induced charge redistribution in the rigid diamond lattice.

Metal-organic frameworks present the most challenging case, with MOFs and extreme MOFs yielding MAEs of $2.57 \pm 1.94$~eV and a roughly linear increase in potential error with charge magnitude. This trend indicates that the dominant source of error in these systems is the magnitude of the net charge redistribution, driven by the large pore volumes and extended coordination environments, rather than a systematic failure of the model architecture. Despite this, even at the most extreme charge states ($|Q| = 20$) the potential $R^2$ remains above 0.99, confirming that the long-range electrostatic structure is faithfully captured.

Taken together, these results show that both models preserve long-range electrostatic information well, but with different strengths: ResNet attains lower potential MAE on several comparatively in-distribution classes, whereas ChargeFlow achieves higher overall potential $R^2$ and clearly stronger downstream performance on the charge-redistribution-heavy organic and diamond-defect stress tests. This distinction is important for applications such as defect formation energy calculations, surface adsorption modeling, and crystal-engineering workflows that depend on robust behavior under distribution shift.

\subsection{Charge-State Extrapolation}

To probe the robustness of ChargeFlow under extrapolation in the charge degree of freedom, we studied prediction error as a function of charge magnitude $|Q|$ for two structurally complex material classes: metal-organic frameworks (MOFs) and organic crystals. The training data contain only integer charges in the range $Q \in \{-3, \ldots, +3\}$; any evaluation at $|Q| > 3$ therefore tests the model's ability to extrapolate beyond the charge range seen during training.

The MOF benchmark provides a moderately sized extrapolation test, whereas the normal/extreme organic-crystal subsets contain only 12 structures each and should therefore be interpreted as a focused stress test rather than a statistically broad benchmark.

\begin{figure}[h!]
    \centering
    \includegraphics[width=0.8\linewidth]{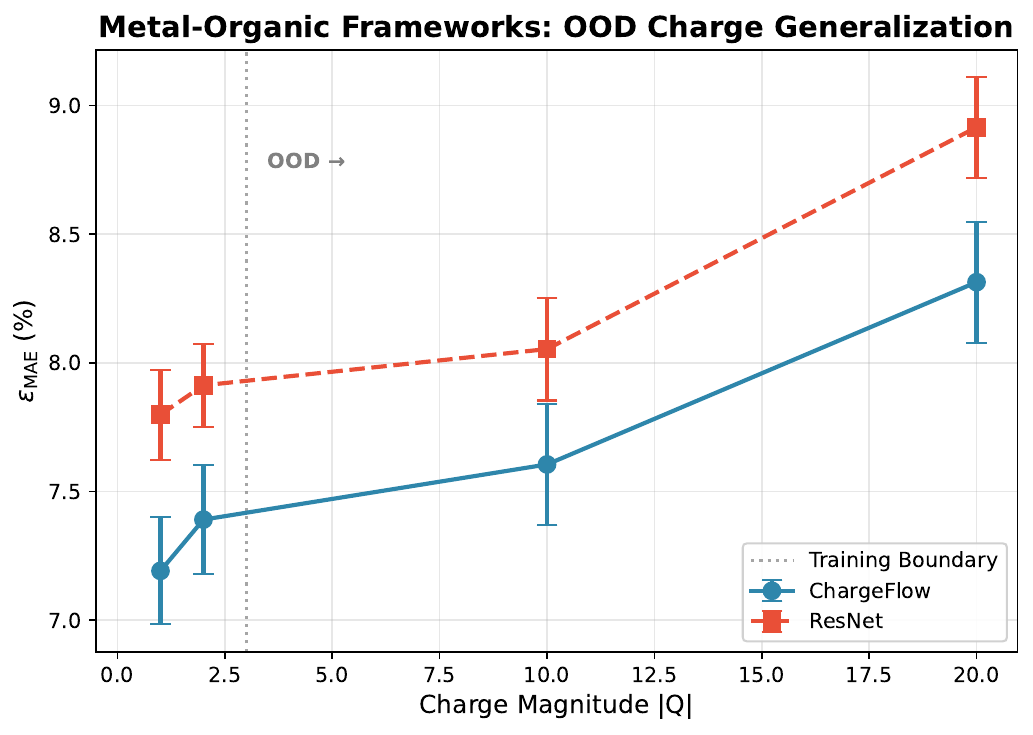}
    \caption{Charge-state extrapolation for metal-organic frameworks. $\varepsilon_{\text{MAE}}$ is plotted as a function of charge magnitude $|Q|$ for ChargeFlow and ResNet. The dashed vertical line marks the training boundary ($|Q| = 3$). ChargeFlow maintains lower error across all charge magnitudes, with the gap widening in the extrapolative regime ($|Q| = 10, 20$). Error bars indicate the standard error of the mean.}
    \label{fig:ood_mofs}
\end{figure}

\textbf{Metal-Organic Frameworks.} Fig.~\ref{fig:ood_mofs} plots $\varepsilon_{\text{MAE}}$ against $|Q|$ for ChargeFlow and the ResNet baseline. Within the training range ($|Q| \leq 3$), ChargeFlow achieves $\varepsilon_{\text{MAE}} \approx 7.2$\% at $|Q| = 1$, rising gently to $\approx 7.4$\% at $|Q| = 2$, while ResNet starts higher at $\approx 7.8$\%. Beyond the training boundary, both models exhibit a monotonic increase in error, but ChargeFlow's error grows more slowly: at $|Q| = 10$ ChargeFlow reaches $\approx 7.6$\% compared to $\approx 8.1$\% for ResNet, and at $|Q| = 20$ the gap widens further ($\approx 8.3$\% versus $\approx 8.9$\%). The widening separation at large $|Q|$ suggests that the flow-matching framework has learned a more transferable model of charge redistribution within the extended pore geometries of MOFs, rather than merely fitting the observed training charges.

\begin{figure}[h!]
    \centering
    \includegraphics[width=0.8\linewidth]{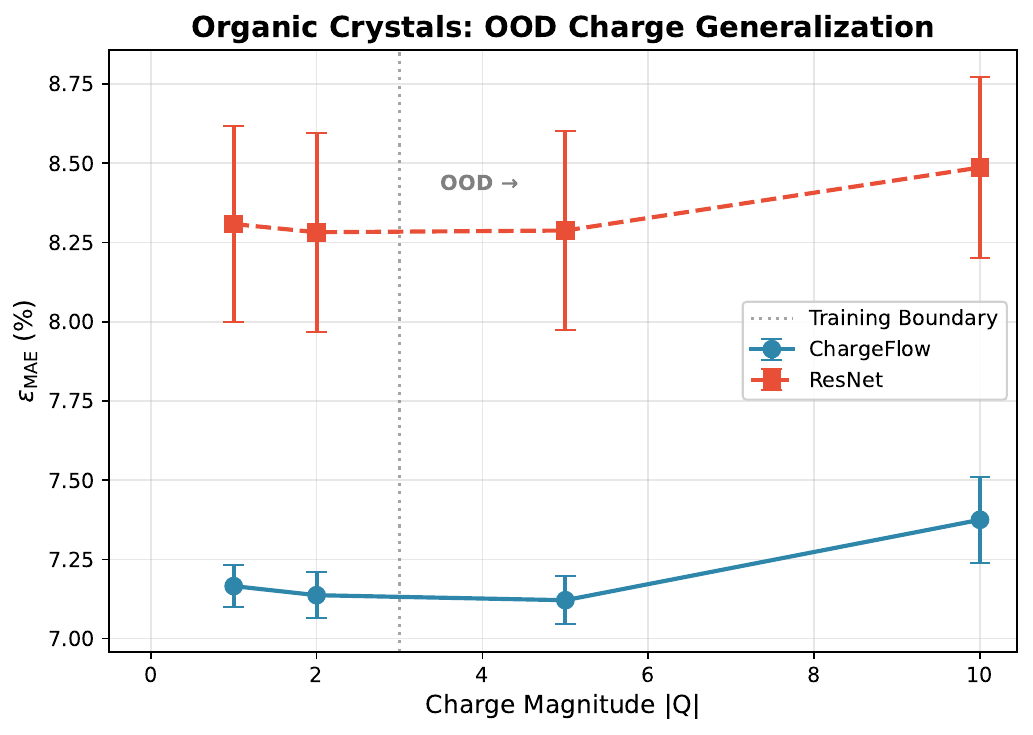}
    \caption{Charge-state extrapolation for organic crystals. ChargeFlow exhibits a remarkably flat error profile ($\varepsilon_{\text{MAE}} \approx 7.1$--$7.4$\%), maintaining a consistent $\sim$1 percentage point advantage over ResNet across both in-range and extrapolative charge states. Error bars indicate the standard error of the mean.}
    \label{fig:ood_organic}
\end{figure}

\textbf{Organic Crystals.} The organic-crystal stress test (Fig.~\ref{fig:ood_organic}) shows a similar trend. ChargeFlow maintains a nearly flat error profile across the available charge range, with $\varepsilon_{\text{MAE}}$ rising only modestly from $\approx 7.15$\% at $|Q| = 1$ to $\approx 7.37$\% at $|Q| = 10$. By contrast, ResNet exhibits a consistently higher error floor of roughly 8.3--8.5\%. Given the small number of structures in this subset, we interpret this result as targeted evidence that ChargeFlow extrapolates charge dependence more smoothly in molecular crystals, rather than as a large-sample benchmark on its own.

These results complement the aggregate statistics in Table~\ref{tab:mae_results} by revealing the \textit{scaling behavior} of error with charge magnitude. The graceful degradation of ChargeFlow's accuracy under extreme charge perturbations supports the hypothesis that the continuous normalizing flow learns a smooth, physically grounded mapping from charge state to electron density, enabling reliable extrapolation far beyond the training distribution.

\subsection{Spatial Localization of Error}

\begin{figure}[h!]
    \centering
    \includegraphics[width=\linewidth]{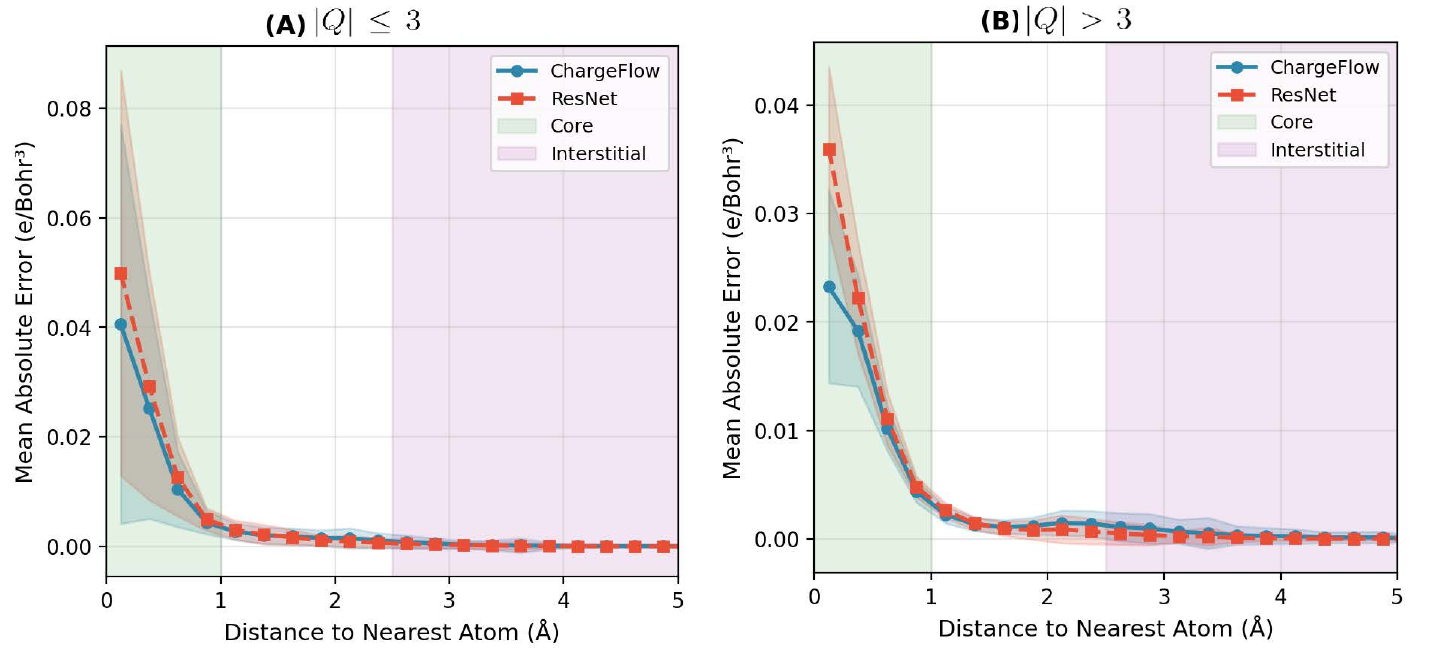}
    \caption{Spatial decomposition of prediction error for ChargeFlow and ResNet. Mean absolute error (e/Bohr$^3$) is plotted as a function of distance to the nearest atom, with shaded bands indicating $\pm 1$ standard deviation across samples. Green and purple backgrounds denote the near-atomic core ($d < 1$~\AA) and interstitial ($d > 2.5$~\AA) regions, respectively. (A)~Within the training charge range ($|Q| \leq 3$), ChargeFlow lowers the peak near-atomic error relative to ResNet. (B)~For extrapolative charge states ($|Q| > 3$), the near-atomic advantage becomes larger, consistent with improved modeling of charge redistribution under stronger perturbations.}
    \label{fig:radial_error}
\end{figure}

To understand where the prediction errors are concentrated in real space, we decomposed the voxel-wise absolute error by the distance from each grid point to its nearest atom. This radial profile separates the near-atomic core region ($d < 1$~\AA), an intermediate bonding region, and the low-density interstitial region ($d > 2.5$~\AA). Figure~\ref{fig:radial_error} compares ChargeFlow and ResNet for systems within the training charge range ($|Q| \leq 3$) and for extrapolative charge states ($|Q| > 3$).

In both regimes, the largest errors occur close to atoms and decay rapidly with distance. ChargeFlow reduces the peak near-atomic error by approximately 20\% within the training charge range and by approximately 30\% beyond it, while remaining comparable to ResNet in the bonding and interstitial regions. This spatial pattern is chemically meaningful: the main advantage of ChargeFlow appears where charge redistribution is largest and where downstream observables such as Bader charges, deformation densities, and electrostatic potentials are most sensitive to density errors.

\subsection{Deformation Density Accuracy}

Another test of chemically relevant density prediction is the deformation density,
\begin{equation}
    \Delta \rho(\mathbf{r}) = \rho(\mathbf{r}) - \rho_{\text{promol}}(\mathbf{r}),
\end{equation}
where $\rho_{\text{promol}}$ is a promolecular superposition of atomic reference densities. Unlike total-density metrics, this removes the dominant near-core atomic contributions that are largely fixed by stoichiometry and isolates the smaller bonding, polarization, and charge-transfer signal. Because ChargeFlow is trained to refine an atomic reference density into the DFT target, accuracy on $\Delta \rho$ directly probes whether the learned flow captures the physically meaningful redistribution of electrons.

\begin{figure}[h!]
    \centering
    \includegraphics[width=\linewidth]{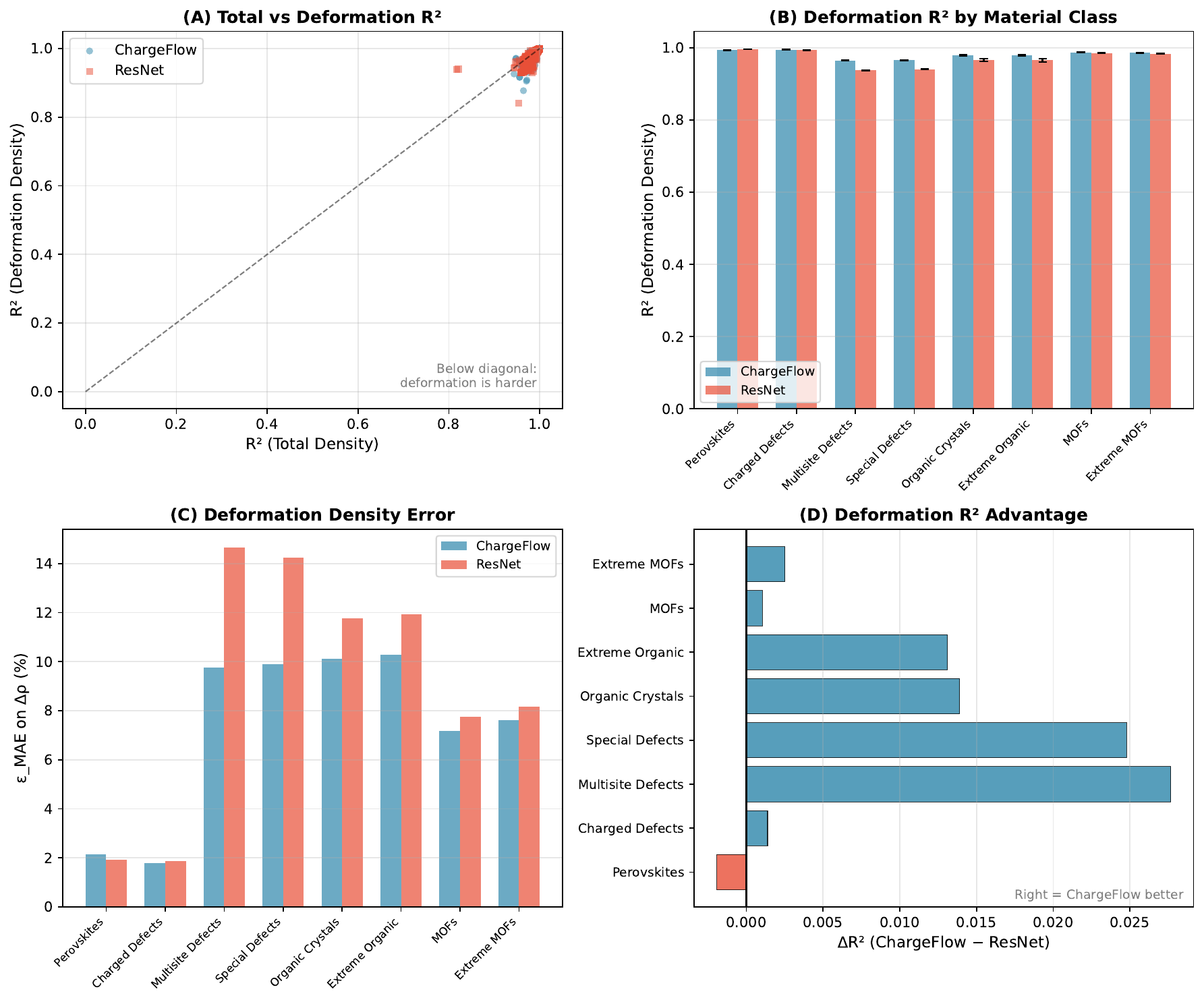}
    \caption{Deformation density analysis comparing ChargeFlow and ResNet. The deformation density $\Delta \rho = \rho_{\text{DFT}} - \rho_{\text{promol}}$ isolates the bonding and charge-transfer contribution after subtracting a promolecular reference. (A)~Scatter of per-structure $R^2$ on total density versus deformation density; points below the diagonal indicate that $\Delta \rho$ is a harder target than the full density. (B)~Mean $R^2$ on $\Delta \rho$ by material class. (C)~Normalized mean absolute error $\varepsilon_{\text{MAE}}$ on $\Delta \rho$ by material class. (D)~Per-class deformation-density advantage, reported as $\Delta R^2 = R^2_{\mathrm{ChargeFlow}} - R^2_{\mathrm{ResNet}}$, where positive values favor ChargeFlow. ChargeFlow outperforms ResNet on 7 of 8 classes, with the largest gains on defective diamonds, organic crystals, and extreme-charge systems.}
    \label{fig:deformation_density}
\end{figure}

We evaluated deformation density fidelity on 1,671 periodic test structures across all eight material classes using the voxel-level coefficient of determination $R^2$ and a normalized mean absolute error $\varepsilon_{\text{MAE}}$ computed on $\Delta \rho$. Results are summarized in Fig.~\ref{fig:deformation_density}. As expected, deformation density is a more difficult target than the full density for both models (Fig.~\ref{fig:deformation_density}A), since the subtraction removes the trivially reconstructible atomic background and leaves only the fine-scale quantum correction. Nevertheless, ChargeFlow retains this signal more faithfully than ResNet: averaged over all structures, ChargeFlow achieves $R^2 = 0.9915$ on $\Delta \rho$ versus 0.9888 for ResNet, while reducing $\varepsilon_{\text{MAE}}$ from 3.62\% to 3.21\%, an 11\% relative improvement. The drop from total-density to deformation-density $R^2$ is also smaller for ChargeFlow (0.0026) than for ResNet (0.0033), indicating that the flow-matching formulation is less reliant on fitting the easy atomic baseline.

The class-wise breakdown in Fig.~\ref{fig:deformation_density}B--D shows that ChargeFlow's advantage is concentrated in the systems where non-local charge redistribution is most important. The largest improvements appear on multisite diamond defects ($\varepsilon_{\text{MAE}} = 9.77$\% vs.\ 14.65\%), special diamond defects (9.91\% vs.\ 14.24\%), organic crystals (10.12\% vs.\ 11.75\%), and extreme organic crystals (10.28\% vs.\ 11.93\%), with consistent gains also observed for MOFs and extreme MOFs. When aggregated over the six most challenging charge-extrapolative and charge-redistribution classes (diamond defects, MOFs, and organics), ChargeFlow reduces the deformation-density error from 9.95\% to 8.23\% and improves $R^2$ from 0.9713 to 0.9804. Perovskites are the main exception: ResNet is slightly better on this well-represented in-distribution class, consistent with the trend already observed in Table~\ref{tab:mae_results}. Overall, these results show that ChargeFlow more accurately captures the chemically meaningful part of the density, namely the redistribution away from a simple atomic superposition toward the bonded DFT ground state.

\subsection{Charge-Response Analysis}

Motivated by recent ML work on learning electron-density response in real space \cite{feng2025efficient}, a key test of whether a model has learned physically meaningful charge-dependent refinement behavior is its ability to predict the \textit{differential density} $\delta\rho = \rho(Q_2) - \rho(Q_1)$, i.e., the change in electron density induced by a change in system charge. Because $\delta\rho$ is a small signal obtained by subtracting two similar density fields, even modest pointwise errors can substantially corrupt the predicted charge response. We quantify the fidelity of the predicted response using the cosine similarity between the ground-truth and predicted $\delta\rho$ vectors (flattened over all voxels), which measures how well the model captures the \textit{spatial pattern} of charge redistribution independently of its overall magnitude.

\begin{figure}[h!]
    \centering
    \includegraphics[width=\linewidth]{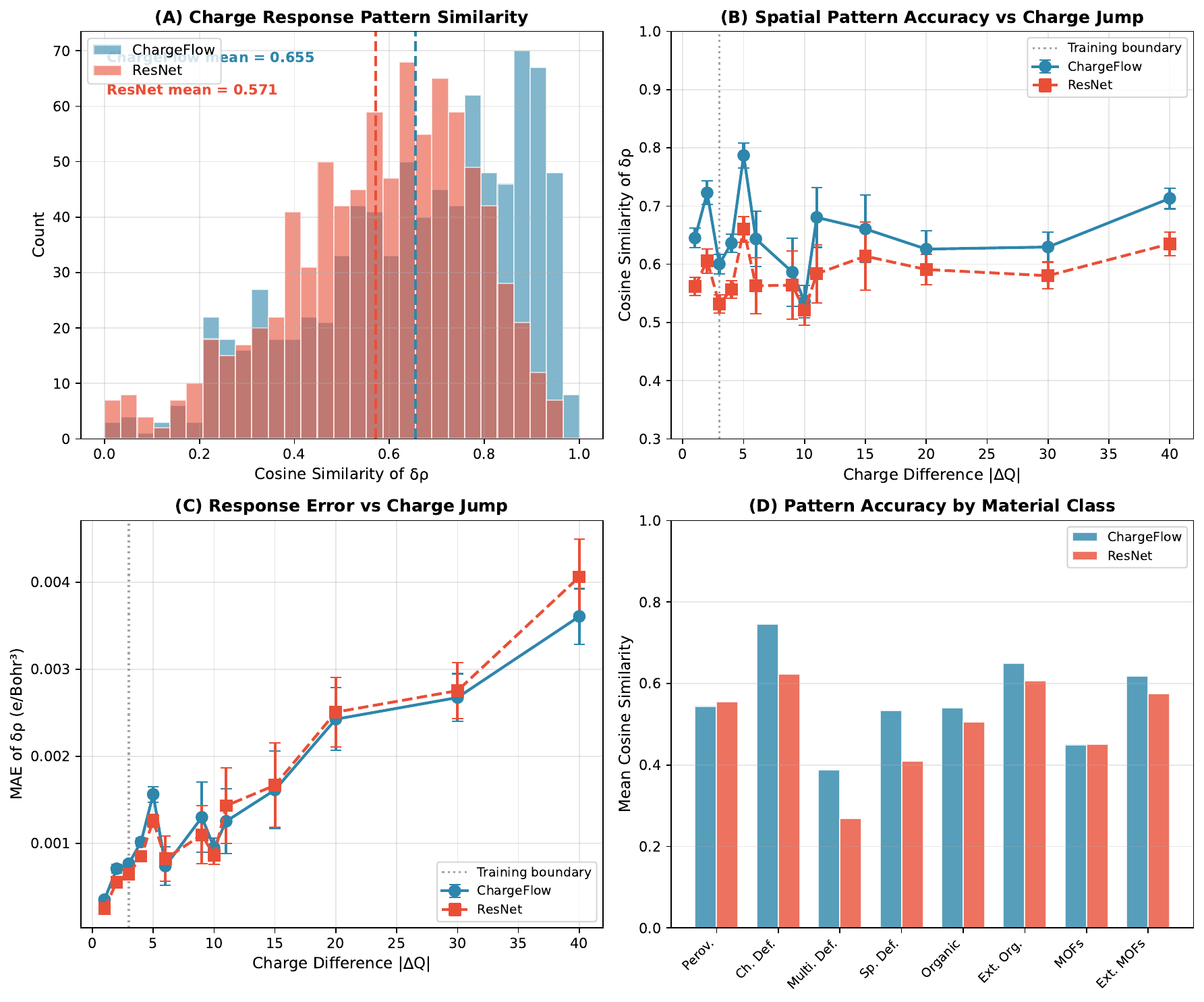}
    \caption{Charge-response analysis comparing ChargeFlow and ResNet across 862 sampled charge-state pairs. (A)~Distribution of cosine similarity between predicted and ground-truth differential density $\delta\rho = \rho(Q_2) - \rho(Q_1)$. (B)~Cosine similarity as a function of charge-jump magnitude $|\Delta Q|$, with the dashed line marking the training boundary ($|\Delta Q| = 3$). (C)~MAE of $\delta\rho$ versus $|\Delta Q|$, showing that ChargeFlow achieves lower error at large charge perturbations. (D)~Mean cosine similarity by material class.}
    \label{fig:charge_response}
\end{figure}

We computed $\delta\rho$ for 862 sampled charge-state pairs across the eight charged test sets used in this analysis, plus matched organic-to-extreme-organic cross-set pairs. Within each test set, the lowest available charge state for a material was used as the reference and paired with higher charges, with at most 500 within-set pairs retained per set. Results are shown in Fig.~\ref{fig:charge_response}.

\textbf{Overall Pattern Accuracy.} ChargeFlow achieves a mean cosine similarity of 0.655 across all 862 pairs, compared to 0.571 for ResNet (Fig.~\ref{fig:charge_response}A), representing a 15\% relative improvement in spatial pattern fidelity. ChargeFlow's advantage is consistent across all charge jump magnitudes $|\Delta Q|$ (Fig.~\ref{fig:charge_response}B), with the gap maintained or widening once the charge jump exceeds the training range ($|\Delta Q| > 3$). At the largest charge jumps tested ($|\Delta Q| = 40$), ChargeFlow achieves a cosine similarity of 0.713 versus 0.635 for ResNet, indicating that the flow-matching framework produces charge-response patterns that remain spatially coherent even under extreme perturbations.

\textbf{Response Error Scaling.} ChargeFlow's main advantage in this analysis is pattern fidelity rather than uniformly lower magnitude error. Averaged over all sampled pairs, ResNet attains a slightly lower mean $\delta\rho$ MAE ($8.0 \times 10^{-4}$ versus $9.2 \times 10^{-4}$~e/Bohr$^3$), largely because many pairs involve relatively small charge jumps. However, the trend reverses in the large-perturbation regime (Fig.~\ref{fig:charge_response}C): at $|\Delta Q| = 40$, ChargeFlow's MAE is $3.6 \times 10^{-3}$~e/Bohr$^3$ compared to $4.1 \times 10^{-3}$~e/Bohr$^3$ for ResNet, a 12\% reduction.

\textbf{Material Class Breakdown.} The per-class cosine similarity (Fig.~\ref{fig:charge_response}D) reveals that ChargeFlow outperforms ResNet on 6 of the 8 material classes displayed in the figure. The largest advantages appear on charged defects (0.746 vs.\ 0.624), special diamond defects (0.534 vs.\ 0.410), and extreme MOFs (0.618 vs.\ 0.575). Perovskites slightly favor ResNet (0.544 vs.\ 0.554), while MOFs are essentially tied (0.450 for both models). These results demonstrate that ChargeFlow produces more physically faithful charge-response functions, particularly for defective and strongly charged systems where the spatial pattern of electron redistribution is complex and non-trivial.

\section{Conclusions}

We have introduced ChargeFlow, a flow-matching model for refining charge-conditioned atomic reference densities into DFT electron densities on native periodic grids. By combining a 3D U-Net velocity field with a direct SAD-to-DFT refinement formulation, the method turns charge-density prediction into a generative transport problem rather than a purely pointwise regression task.

Across a chemically heterogeneous external benchmark, ChargeFlow is not uniformly best on every in-distribution class, but it is consistently strongest on the tasks that motivated the method: nonlocal charge redistribution, deformation-density fidelity, and extrapolation beyond the training charge range. The downstream observables sharpen this conclusion. Relative to a ResNet baseline, ChargeFlow yields more robust Bader partitioning (1671/1671 successful structures versus 1590/1671), higher atom-level Bader fidelity, more accurate deformation densities, and more faithful charge-response patterns, while maintaining high-fidelity electrostatic potentials. ChargeFlow also offers an approximate three-order-of-magnitude speed-up over conventional DFT for inference, making rapid charge-state exploration practical.

Future work should focus on parent-structure-aware training splits, broader charged benchmarks, extension to spin density and related observables, and tighter integration of density refinement into electronic-structure workflows such as defect screening and self-consistent initialization.

\section*{Associated Content}
\textbf{Supporting Information.} The Supporting Information includes block-level architecture details, additional notes on the development split and evaluation protocol, training-corpus statistics, representative qualitative slices, ResNet-specific downstream figures, and additional global diagnostics on charge scaling and electron conservation.

\section*{Data and Code Availability}
The dataset release supporting this work is publicly available on Zenodo at \url{https://zenodo.org/records/19211405} (DOI: \url{https://doi.org/10.5281/zenodo.19211405}). The ChargeFlow training, inference, and analysis code is publicly available at \url{https://github.com/ngminhtri0394/chargeflow-electron-density}.

\end{document}